\documentclass[mathleft
]{an}
\usepackage{graphicx}
\usepackage{txfonts}

\newcommand{\vsini}{$v \sin i$}

\newcommand{\logg}{$\log(g)$}

\newcommand{\logteff}{$\log(T_{\mathrm{eff}})$}
\newcommand{\logl}{$\log(L/L_{\odot})$}

\usepackage{times}
\begin{document}

\Pagespan{789}{}
\Yearpublication{2006}%
\Yearsubmission{2005}%
\Month{11}%
\Volume{999}%
\Issue{88}%

\title{Discovery of magnetic fields in three He variable Bp stars with He and Si spots\thanks{Based
on observations collected at the European Southern Observatory, Paranal, Chile
(ESO programmes 71.D-0308(A), 072.D-0377(A), and 073.D-0466(A)).}}

\author{{M. Briquet\inst{1}\fnmsep\thanks{Postdoctoral fellow of the Fund for Scientific Research, Flanders.}\fnmsep\thanks{Corresponding author:
  \email{maryline@ster.kuleuven.be}}}
\and  S. Hubrig\inst{2}
\and M. Sch\"oller\inst{2}
\and P. De Cat\inst{3}
}
\titlerunning{Discovery of magnetic fields in three He variable Bp stars with He and Si spots}
\authorrunning{Briquet et al.}
\institute{
Instituut voor Sterrenkunde, Katholieke Universiteit Leuven, Celestijnenlaan 200 B, B-3001 Leuven, Belgium
\and 
European Southern Observatory, Casilla 19001, Santiago 19, Chile
\and 
 Koninklijke Sterrenwacht van Belgi\"e, Ringlaan 3, B-1180 Brussel, Belgium}

\received{}
\accepted{}
\publonline{later}


\abstract{%
It is essential for the understanding of stellar structure models of high mass stars to explain why constant stars, non-pulsating chemically peculiar hot Bp stars and pulsating stars 
co-exist in the slowly pulsating B stars and $\beta$\,Cephei 
instability strips.
We have conducted a search for magnetic fields in the four Bp stars
HD\,55522, HD\,105382, HD\,131120, and HD\,138769 which previously have been wrongly identified as 
slowly pulsating B stars. A recent study of these stars using the Doppler Imaging technique 
revealed that the elements He and Si are inhomogeneously distributed 
on the stellar surface, causing the periodic variability.
Using FORS\,1 in spectropolarimetric mode at the VLT, we have acquired circular polarisation
spectra 
to test the presence of a magnetic field in these stars.
A variable magnetic field is clearly detected in HD\,55522 and HD\,105382, but no evidence
for the existence of a magnetic field was found in HD\,131120.
The presence of a magnetic field in HD\,138769 is suggested by one measurement at 3\,$\sigma$ level. 
We discuss the occurence of magnetic B stars among the confirmed pulsating B stars and find strong magnetic fields of order kG and oscillations to be mutually exclusive. }

\maketitle

\section{Introduction}
\begin{table*}
\caption{
The mean longitudinal field measurements for our sample
of Bp stars observed
with FORS\,1 in the frame of our ESO service programs
71.D-0308, 072.D-0377, and 073.D-0466.
In the first four columns we give the HD number, another identifier,
the $V$ magnitude and the spectral type.
In Columns~5 and 6 we present the modified Julian date of the middle of the exposures
and the corresponding measured mean longitudinal magnetic field $\left<B_{\mathrm l}\right>$.
For each star
we give the rms longitudinal magnetic field
and the reduced $\chi^2$ for all measurements in Columns~7 and 8.
}
\label{table1}
\begin{center}
\begin{tabular}{rlrllrrr}
\hline
\multicolumn{1}{c}{\raisebox{2mm}{\rule{0mm}{2mm}}HD} &
\multicolumn{1}{c}{Other} &
\multicolumn{1}{c}{$V$} &
\multicolumn{1}{c}{Sp.\ Type} &
\multicolumn{1}{c}{MJD} &
\multicolumn{1}{c}{$\left<B_{\mathrm l}\right>$} &
\multicolumn{1}{c}{$\overline{\left< B_l \right>}$} &
\multicolumn{1}{c}{$\chi^2/n$} \\
\multicolumn{1}{c}{} &
\multicolumn{1}{c}{Identifier} &
\multicolumn{1}{c}{} &
\multicolumn{1}{c}{} &
\multicolumn{1}{c}{} &
\multicolumn{1}{c}{[G]} &
\multicolumn{1}{c}{[G]} &
\multicolumn{1}{c}{} \\
\hline
55522   &  HR2718  &  5.9  &  B2IV/V       &  52999.190  & 38$\pm$\,~73 & 518 & 65.1 \\
 & & & &  52999.227  & 39$\pm$234 & & \\
 & & & &  53000.053  & 873$\pm$\,~66 & & \\
 & & & &  53275.295	 & 554$\pm$\,~60 & & \\
105382  &  HR4618  &  4.4  &  B6IIIp He-weak  &  53011.195  & $-$923$\pm$\,~86 & 751 & 105.6 \\
 & & & &  53015.247  & $-$431$\pm$109 & & \\
 & & & &  53144.003	 & 840$\pm$\,~58 & & \\
 & & & &  53224.989	 & $-$715$\pm$\,~79 & & \\
131120  &  HR5543  &  5.0  &  B7p He-weak   &  52824.158  & $-$228$\pm$110 & 130 & 2.3 \\
 & & & &  53020.353  & $-$137$\pm$\,~74 & & \\
 & & & &  53030.366  & 63$\pm$\,~69 & & \\
 & & & &  53225.027	 & 92$\pm$\,~57 & & \\
 & & & &  53234.102	 & $-$39$\pm$\,~72 & & \\
138769  &  HR5781  &  4.5  &  B3IVp He-weak       &  52904.027  & $-$16$\pm$\,~58 & 225 & 5.9 \\
 & & & &  52908.022  & $-$260$\pm$\,~84 & & \\
 & & & &  53234.120	 & $-$289$\pm$101 & & \\
\hline
\end{tabular}
\end{center}
\end{table*}

In the framework of a long-term monitoring project dedicated to the study of 
seismic models for a large sample of slowly pulsating B (SPB) stars
(Aerts et~al.\ \cite{aerts}, Mathias et~al.\ \cite{mathias}) four variable B-type
stars -- HD\,55522, HD 105382, HD\,131120, and HD\,138769 -- have been identified as 
chemically peculiar He -variable stars (Briquet et~al.\ \cite{briquet3}). 
All available data sets revealed that these stars are monoperiodic variables. A comparison of moment variations of silicon 
and helium lines allowed to conclude that the observed variability 
of these four stars must be attributed to an inhomogeneous distribution of 
chemical elements on the stellar surface.
All four stars are 
moderate rotators with \vsini{} $\in$ [70,86] km/s. Using high 
resolution CAT/CES spectra, 
Briquet et~al.\ (\cite{briquet3}) produced abundance maps for both silicon and helium on the 
stellar surface using the Doppler Imaging technique. 
Currently, these maps are the only existing Doppler images generated for 
hot B-type stars with an effective temperature as high as 18,000\,K. 

All previous studies of chemically peculiar Ap and Bp stars with inhomogeneous distributions
of elements on the stellar surface show that these stars are also known to have variable 
magnetic fields, generally diagnosed through mean longitudinal magnetic 
field, mean magnetic field
modulus or net broadband linear polarisation measurements. When these 
observations are interpreted within the framework of the oblique rotator 
model, we obtain a self-consistent picture of a star whose atmosphere 
contains a heterogeneous distribution of chemical elements, and is permeated 
by a static, quasi-dipolar magnetic field. As the star rotates, we observe 
the magnetic field and surface
abundance distribution from various aspects, resulting in variability of the
measured magnetic field and line strength.
The peculiarities in the Ap and Bp stars are believed to result from an upward 
selective diffusion
of some elements, for which the radiative force exceeds the downwards gravitational settling of other elements. Such a 
separation produces apparent overabundances in the floating elements and
underabundances of the sinking ones in the atmospheres of chemically peculiar
stars (e.g., Michaud \ \cite{michaud2}; Michaud et~al.\ \cite{michaud1}).

Four magnetic field observations have been obtained by Borra et~al.\ (\cite{borra})
for HD\,131120 in 1981.
However, they showed no sign of the presence of a magnetic field. 
No measurements of magnetic fields for HD\,55522, HD 105382, and HD\,138769
have been reported in the literature.
In this paper, we present the first magnetic observations of all four stars which we gathered 
with FORS\,1 at the VLT in the last years and discuss the magnetic data in view of 
the He and Si distributions derived in Briquet et~al.\ (\cite{briquet3}).

\section{Observations}

The observations reported here have been carried out
at the European Southern Observatory with FORS\,1
(FOcal Reducer low dispersion Spectrograph)
mounted on the 8-m Melipal telescope of the VLT.
This multi-mode instrument is equipped with polarisation analyzing optics comprising super-achromatic half-wave and quarter-wave phase  retarder plates,
and a Wollaston prism with a beam divergence of 22$\arcsec$ in standard resolution mode.
We used the GRISM\,600B in the wavelength range 3480--5890 $\AA$ 
to cover all hydrogen Balmer lines from H$\beta$ to the Balmer jump.
The spectral resolution of the FORS\,1 spectra taken with this setting was $R\sim2000$. 
The determination of the mean longitudinal fields
using FORS\,1 is described in detail in Hubrig et~al.\ (\cite{hubrig04}).
While the accuracy of the longitudinal magnetic field determination using metal lines
strongly depends on the width of the spectral lines used in the analysis,  the advantage of 
using  FORS\,1 is especially obvious in fast rotating stars with large \vsini{} values where for the 
measurement of polarisation exclusively intrinsically broad 
hydrogen Balmer lines are used.

For each star we usually took four to eight continuous series of two exposures
with the retarder waveplate oriented at different angles.
The spectropolarimetric capability of the FORS\,1 instrument
in combination with the large light collecting power
of the VLT allows us to achieve a S/N ratio up to a few thousands per pixel
in the one--dimensional spectrum, as required to detect low
polarisation signatures produced in the spectral lines
by longitudinal magnetic fields of the order of hundred Gauss and less.

\begin{figure*}  
\centering
\includegraphics[width=0.24\textwidth]{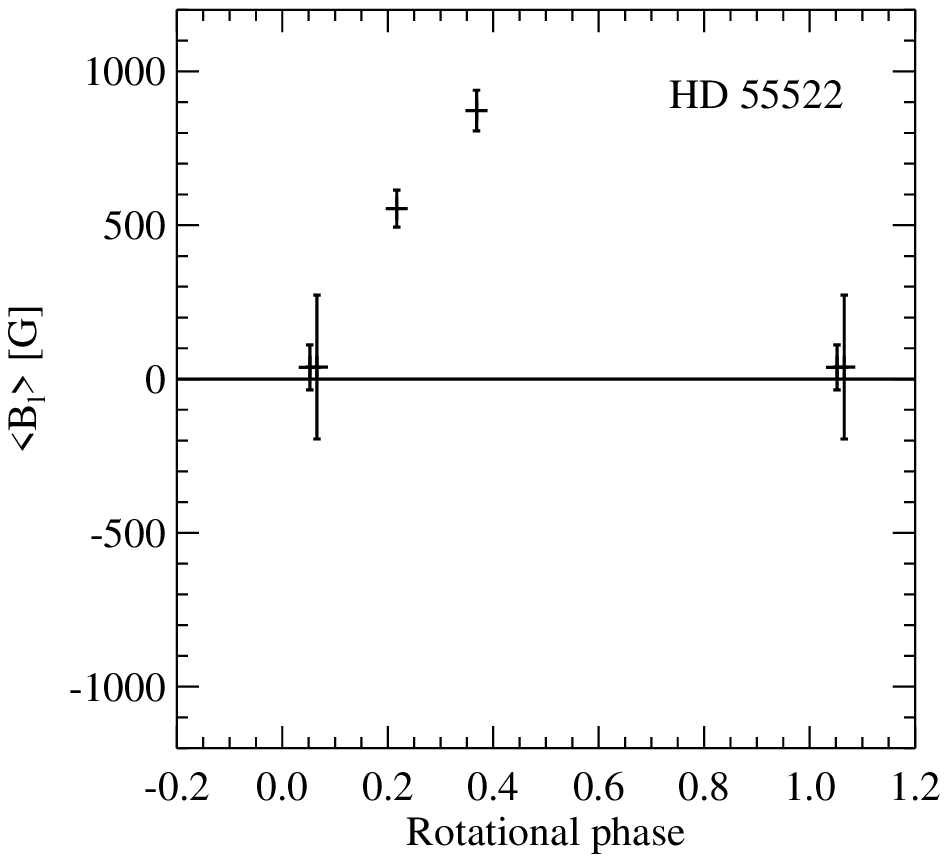}
\includegraphics[width=0.24\textwidth]{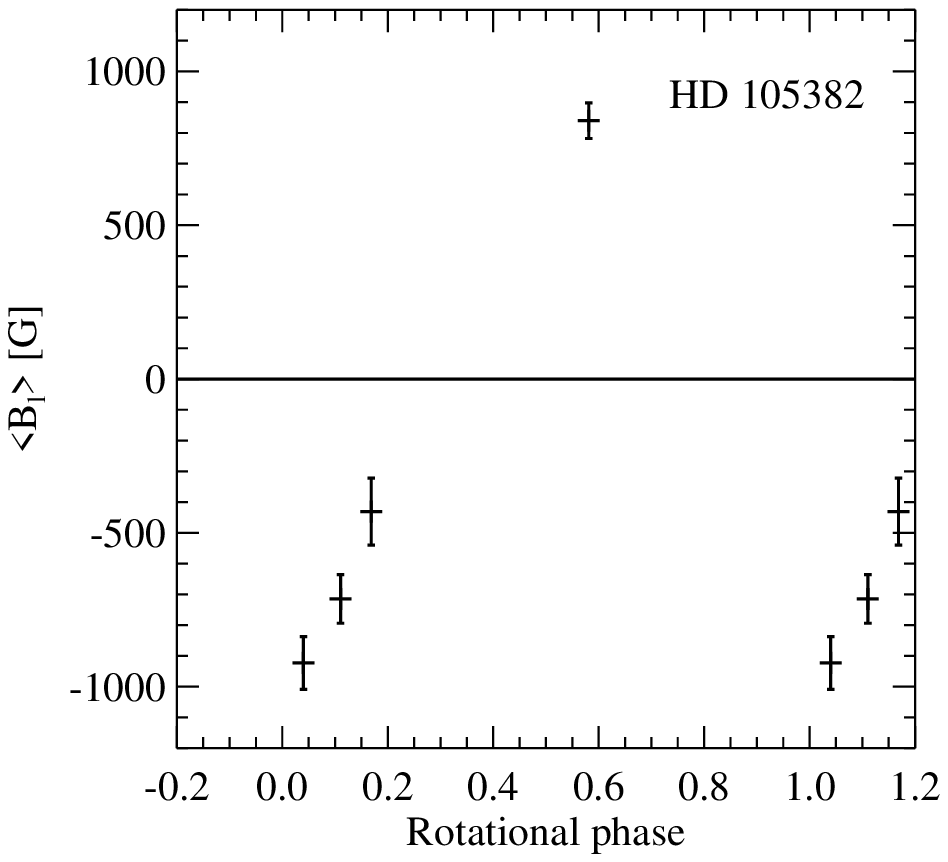}
\includegraphics[width=0.24\textwidth]{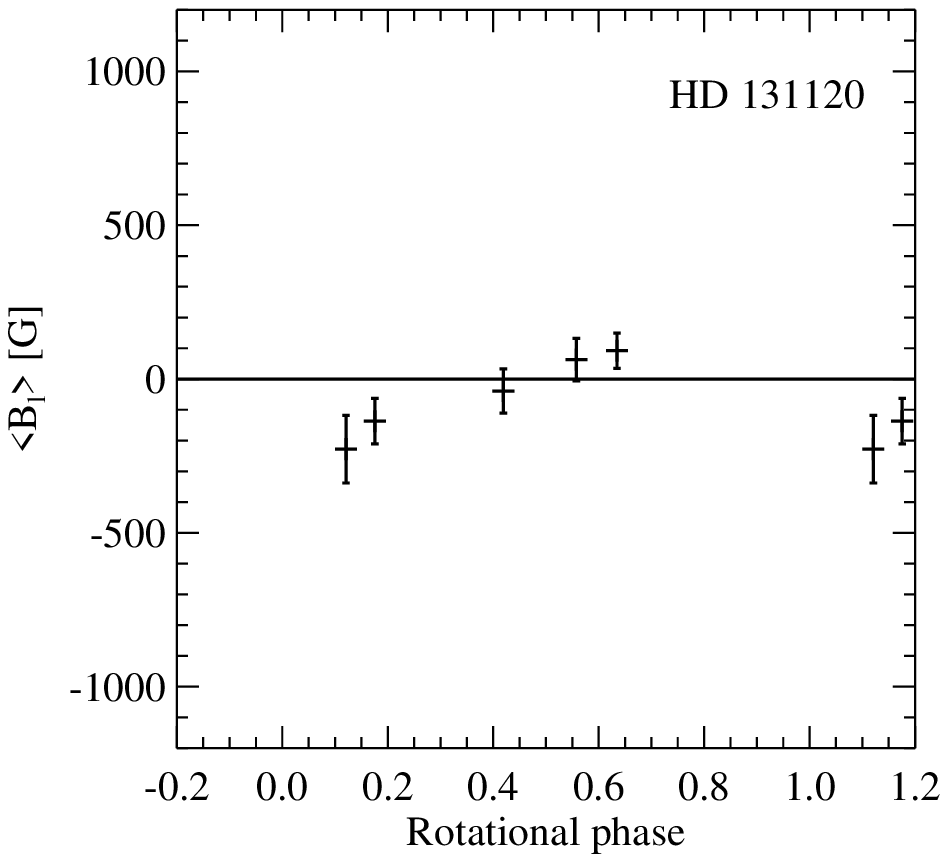}
\includegraphics[width=0.24\textwidth]{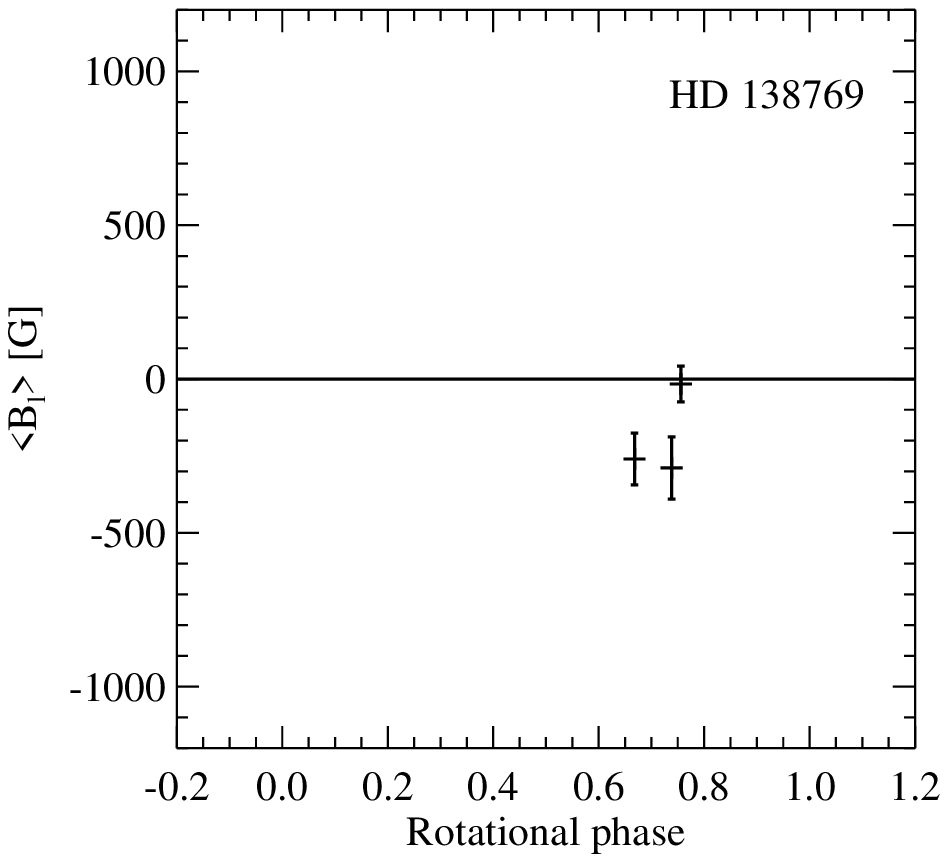}
\caption{The mean longitudinal magnetic field variability over the stellar rotation cycle
for HD\,55522, HD\,105382, HD\,131120, and HD\,138769.
}   
\label{fig:mag}  
\end{figure*}  

The mean longitudinal magnetic field is the average over the stellar hemisphere
visible at the time of observation of the component of the magnetic field parallel
to the line of sight, weighted by the local emergent spectral line intensity.
It is diagnosed from the slope of a linear regression of $V/I$ versus the quantity
$-g_{\rm eff} \Delta\lambda_z \lambda^2 \frac{1}{I} \frac{{\mathrm d}I}{{\mathrm d}\lambda} \left<B_l\right > 
+ V_0/I_0$ (Bagnulo et~al.\ \cite{bagnulo}).
The study of a large sample of magnetic and non-magnetic Ap and Bp stars already proved that this regression technique is very robust so that stars with detections with $\left<B_l\right> > 3$\,$\sigma$ possess magnetic fields.

Individual magnetic field measurements are given in Table~\ref{table1}.
In the first four columns we give the HD number, another identifier, the visual magnitude and the spectral
type. The modified Julian date of the middle of the exposures and
the measured mean longitudinal magnetic field $\left<B_{\mathrm l}\right>$
are presented in Columns 5 and 6 respectively.
The rms longitudinal magnetic field is given in Column~7.
It is computed from all $n$ measurements according to:
\begin{equation}
\overline{\left< B_l \right>} = \left( \frac{1}{n} \sum^{n}_{i=1} \left< B_l \right> ^2_i \right)^{1/2}.
\label{eqn1}
\end{equation}

The reduced $\chi^2$ for 
these measurements are presented in Column 8, following:
\begin{equation}
\chi^2/n = \frac{1}{n} \sum_{i=1}^n \left( \frac{\left< B_l \right>_i}{\sigma_i} \right)^2
.
\end{equation}

Magnetic observations of all four stars phased 
with the rotation periods obtained by Briquet 
et~al.\ (\cite{briquet3}) are presented in Fig.~\ref{fig:mag}.

\section{Discussion}\label{sec3}

The fundamental parameters of 
the studied stars are presented in Table~\ref{table2}.
The effective temperature \logteff{} and the surface gravity \logg{} (Columns~2 and 3 in Table~\ref{table2}) were derived in Briquet et~al.\ (\cite{briquet3}). 
To determine other 
stellar parameters, a grid of main-sequence models has been used,
which was calculated with 
the Code Li\'egeois d'\'Evolution Stellaire
(version 18.2, written by R.\ Scuflaire), assuming solar composition.
For a detailed description, see ``grid 2'' in De Cat et~al.\ (\cite{DeCat2006}).
The mass $M$, the radius $R$, the luminosity \logl, and the age of 
the star expressed as a fraction of its total main-sequence life
$f$ are presented 
in Columns~4 to 7 in Table~\ref{table2}. 
The projected rotational velocity values (Column~8) in Table~\ref{table2} have been taken
from Briquet et~al.\ (\cite{briquet3}).
To determine 
the rotational periods (Column 9) presented in the last column of Table~\ref{table2},
Hipparcos photometry, multi-colour Geneva photometry
and high-resolution CES spectra have been used.

\begin{table*}
\caption{
Fundamental parameters for the objects in our sample.
In the first column we give the HD number.
In the following five columns we list the logarithm of
the effective temperature, the logarithm of the surface gravity, mass,
stellar radius, and the logarithm of the stellar luminosity.
The final three columns give the fraction of the main sequence lifetime
for each individual star, its \vsini{} and the rotational period.
\label{tab:par}
}
\label{table2}
\begin{center}
\begin{tabular}{rrrrrcrcr}
\hline
\multicolumn{1}{c}{HD} &
\multicolumn{1}{c}{\logteff{}} &
\multicolumn{1}{c}{\logg{}} &
\multicolumn{1}{c}{$M/M_\odot$} &
\multicolumn{1}{c}{$R/R_\odot$} &
\multicolumn{1}{c}{\logl{}} &
\multicolumn{1}{c}{f} &
\multicolumn{1}{c}{\vsini{}} &
\multicolumn{1}{c}{P$_{\rm rot}$} \\
\multicolumn{1}{c}{} &
\multicolumn{1}{c}{} &
\multicolumn{1}{c}{} &
\multicolumn{1}{c}{} &
\multicolumn{1}{c}{} &
\multicolumn{1}{c}{} &
\multicolumn{1}{c}{} &
\multicolumn{1}{c}{[km/s]} &
\multicolumn{1}{c}{[d]} \\
\hline
 55522 & 4.241$\pm$0.020 & 4.15$\pm$0.20 & 5.5$\pm$0.9 & 3.3$\pm$1.1 & 3.0$\pm$0.3 & 0.46$\pm$0.33 & 70$\pm$5 & 2.729$\pm$0.001 \\
105382 & 4.241$\pm$0.020 & 4.18$\pm$0.20 & 5.4$\pm$0.9 & 3.3$\pm$1.0 & 2.9$\pm$0.3 & 0.43$\pm$0.32 & 75$\pm$5 & 1.295$\pm$0.001 \\
131120 & 4.261$\pm$0.020 & 4.10$\pm$0.20 & 6.1$\pm$1.1 & 3.7$\pm$1.2 & 3.1$\pm$0.3 & 0.53$\pm$0.30 & 86$\pm$5 & 1.569$\pm$0.001 \\
138769 & 4.243$\pm$0.020 & 4.22$\pm$0.20 & 5.4$\pm$0.8 & 3.2$\pm$0.8 & 2.9$\pm$0.3 & 0.38$\pm$0.30 & 85$\pm$5 & 2.089$\pm$0.001 \\
\hline
\end{tabular}
\end{center}
\end{table*}

Our measurements show a clear variability of the mean longitudinal magnetic field 
over the rotational period on the surface 
of HD\,55522 and HD\,105382 (Fig.~\ref{fig:mag}). The abundance maps for He and Si 
surface distribution obtained by Briquet et~al.\ (\cite{briquet3}) for HD\,55522
revealed depleted silicon regions along the equator 
and enhanced silicon regions located close to the rotational poles.
Around phase 0.75, a strong helium spot is visible on the equator surrounded
by a very depleted region.
If we assume that $\left<B_l\right>$ varies as a sine function with the stellar rotation period,
both  extrema of the magnetic field would correspond to rotational phases where 
He and Si are found depleted. 

The maps for HD\,105382 showed two strongly overabundant He spots just above the equator,
which are visible around phases 0.08 and 0.67. A less strong overabundant third He spot  
became visible at phase 0.5.
The regions with depleted helium showed enhanced silicon and vice versa.
One overabundant He spot (or underabundant Si spot)
corresponds to the negative extremum of the measured $\left<B_l\right>$ while the other overabundant 
He spot (or underabundant Si spot) coincides with the positive magnetic field extremum,
if we assume that the mean longitudinal magnetic field varies sinusoidally with the 
stellar rotation period.

HD\,131120 showed two overabundant He spots close to the equator,
which have been visible in phases 0.2 and 0.5.
One underabundant He spot was visible around the phase 0.75.
The presence of a magnetic field was not confirmed  by Borra et~al.\ (\cite{borra}).
Also none of our five magnetic field measurements is 
significant at 3\,$\sigma$ level. 
Since in general the presence of chemical inhomogeneities on the surface of 
Ap and Bp stars is closely 
connected with the presence of magnetic fields in their atmospheres,
we suggest that HD 131120 is likely a very weak magnetic star with a mean longitudinal 
magnetic field below one hundred Gauss.

\begin{figure}  
\centering
\includegraphics[height=0.45\textwidth,angle=270]{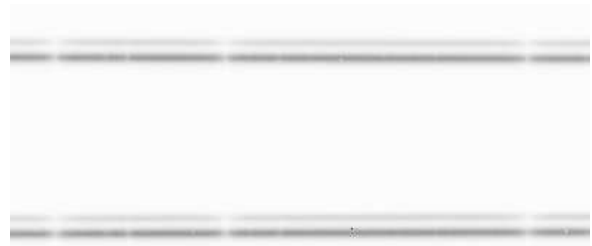}
\caption{FORS\,1 spectra of  HD\,138769 taken in right- and left-hand circular 
polarized light from H$\epsilon$ to H$\gamma$.
The presence of the close visual companion is clearly visible.
}
\label{fig:binary}  
\end{figure}  

For HD\,138769,  Briquet et~al.\ (\cite{briquet3}) found an enhanced Si spot 
located at the equator which is visible around phase 0.5 whereas
a depleted Si region appeared close to the pole.
The existence of the polar spot was considered as less secure.
As for helium, no abundance distribution could be derived
because the deviation between observed and calculated
profiles for the best maps was still very large. The star HD\,138769 is a close visual double star with a separation of 2\farcs{}19$\pm$0\farcs{}03 (e.g.\ Hurly \& Warner\ \cite{hurly}). In one of our FORS\,1 spectra we noticed the presence of the close visual component at a separation of $\sim$1\farcs{}5 (Fig.~\ref{fig:binary}). However, in case of worse seeing the visual components are not separated, and thus the observed spectra are contaminated. Only three magnetic field measurements have been obtained with FORS\,1, and among them only the measurement at rotational phase 0.67 is formally significant at 3\,$\sigma$ level.

\section{Conclusions}
In summary, we conclude that the magnetic fields are clearly detected in  HD\,55522 and HD\,105382,
whereas the presence of a magnetic field in HD 131120 and HD 138769 still needs confirmation by 
more accurate spectropolarimetric observations. Since the measured longitudinal 
magnetic fields in HD\,55522 and HD\,105382 are rather strong, these stars seem to be excellent 
candidates to study their magnetic field geometry based on a larger number of magnetic field measurements.

\begin{figure}  
\centering
\includegraphics[height=0.48\textwidth,angle=270]{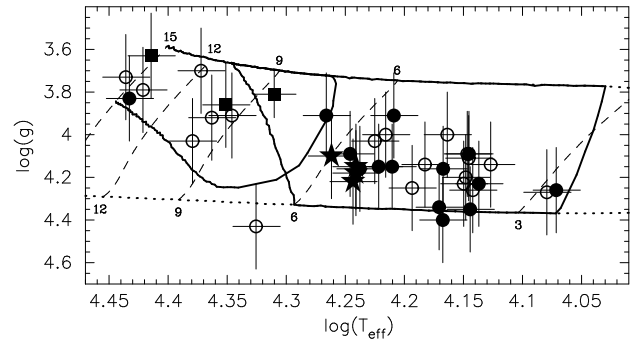}
\caption{
The position of the four studied Bp stars in the H-R diagram is indicated
by stars.
We also show by circles the position of $\beta$ Cephei and SPB stars for which Hubrig et~al. (\cite{hubrig06a}) searched for the presence of a magnetic field. In addition we represent by squares the three magnetic B-type pulsators discovered by Neiner et~al.\ (2003a,b) and Donati et~al.\ (\cite{Donati2001}). Filled circles and squares correspond to stars with detected magnetic fields. The full lines represent the boundary of the theoretical instability strips
for modes with a frequency between 0.2 and 30\,$\rm{d^{-1}}$
and $\ell \le 3$ computed for the main sequence models with
$M \in \left[ 2,15 \right] M_{\odot}$ in ``grid 2'' of De Cat et~al.\ (\cite{DeCat2006})
for which Z=0.015, X=0.71, $\alpha _{\rm conv}$=1.75, $\alpha _{\rm over}$=0.0,
and the standard metal mixture of Asplund et~al.\ (\cite{Asplund2005}).
The lower and upper dotted lines show the ZAMS and TAMS, respectively.
The dashed lines denote evolution tracks for stars with $M$=15, 12, 9, 6,
and 3\,$M_{\odot}$.
}
\label{fig:hr}  
\end{figure}

The position of the four studied He variable Bp stars in the H-R diagram is shown in Fig.~\ref{fig:hr}.
In the same diagram we also present the distribution of 
SPB and $\beta$\,Cephei stars 
with studied magnetic fields (Hubrig et~al.\ \cite{hubrig06a}; Hubrig et~al.\ \cite{hubrig06b}; Neiner et~al.\ \cite{Neiner2003a}; Neiner et~al.\ \cite{Neiner2003b}; Donati et~al.\ \cite{Donati2001}).
Filled circles and squares correspond to the pulsating stars with detected magnetic fields.
Non-pulsating chemically peculiar hot Bp stars and pulsating stars evidently
co-exist in the SPB and $\beta$\,Cephei 
instability strips. It is especially intriguing that the magnetic fields 
of hot Bp stars either do not show any detectable variations or vary with periods
close to one day, 
which is of the order of the pulsation and rotation period range of SPB stars
(Bohlender et~al.\ \cite{Bohlender1987}; Matthews \& Bohlender \cite{Matthews1991}).
The measured magnetic fields in 14 SPB stars and in 3 stars of $\beta$\,Cephei type, however, demonstrate that their fields are rather weak in comparison to the kG fields detected in magnetic Bp stars. This seems to indicate that very strong magnetic fields are not co-existent with oscillations, or stars with stronger magnetic fields have much lower pulsation amplitudes. Such an observational evidence has recently been mentioned for rapidly oscillating stars by Kurtz et al.\ (\cite{kurtz}). We note however that the pulsation mechanism is very different in the two groups of stars.

Clearly, the presently available observational results are still marginal. Further studies of magnetic fields in hot B stars, both pulsating and non-pulsating, are necessary to provide important information on the magnetic field geometry in these stars to test theoretical predictions related to the origin of magnetic fields (e.g., Mullan \& MacDonald\ \cite{mullan}; Spruit\ \cite{spruit}).


\begin{thebibliography}{}
\bibitem[1999]{aerts} Aerts, C., De Cat, P., Peeters, E., et al., 1999, A\&A, 343, 872

\bibitem[2005]{Asplund2005} Asplund, M., Grevesse, N., \& Sauval, A.J., 2005, in ASP Conf.\ Ser.\ Vol.\ 336,
Cosmic Abundances as Records of Stellar Evolution and Nucleosynthesis, eds.\ T.G.\ Barnes III \& F.N. Bash, 25

\bibitem[2002]{bagnulo} Bagnulo, S., Szeifert, T., Wade, G.A., et al. 2002, A\&A, 389, 191

\bibitem[1987]{Bohlender1987} Bohlender, D.A., Landstreet, J.D., Brown, D.N., \& Thompson, I.B., 1987, ApJ, 323, 325

\bibitem[1983]{borra} Borra, E.F., Landstreet, J.D., Thompson, I., 1983, ApJS, 53, 151



\bibitem[2004]{briquet3} Briquet, M., Aerts, C., L\"uftinger, T., et al., 2004, A\&A, 413, 273


\bibitem[2006]{DeCat2006} De Cat, P., Briquet, M., Aerts, C., et al., 2006, CoAst, 147, 48

\bibitem[2001]{Donati2001} Donati, J.-F., Wade, G.A., Babel, J., et al., 2001, MNRAS, 326, 1265

\bibitem[2006a]{hubrig06a} Hubrig, S., Briquet, M., Sch\"oller, M., et al., 2006a, MNRAS, 369, 61

\bibitem[2006b]{hubrig06b} Hubrig, S., North P., Sch\"oller, M., Mathys, G. 2006b, AN, 327, 289

\bibitem[2004]{hubrig04} Hubrig, S., Szeifert, T., Sch\"oller, M., et al., 2004, A\&A, 415, 661

\bibitem[1983]{hurly} Hurly \& Warner, 1983, MNRAS, 202, 761

\bibitem[2006]{kurtz} Kurtz, D.W., Elkin, V.G., Cuhna, M.S. et al., submitted to MNRAS


\bibitem[2001]{mathias} Mathias, P., Aerts, C., Briquet M., et al., 2001, A\&A, 379, 905

\bibitem[1991]{Matthews1991} Matthews, J.M., \& Bohlender, D.A., 1991, A\&A, 243, 148

\bibitem[1970]{michaud2} Michaud, G., 1970, ApJ, 160, 641

\bibitem[1981]{michaud1} Michaud, G., Megessier, C., \& Charland, Y., 1981, A\&A, 103, 244

\bibitem[2005]{mullan} Mullan, D.J., MacDonald, James, 2005, MNRAS, 356, 1139

\bibitem[2003a]{Neiner2003a} Neiner, C., Geers, V.C., Henrichs, H.F., et al., 2003a, A\&A, 406, 1019

\bibitem[2003b]{Neiner2003b} Neiner, C., Henrichs, H.F., Floquet, M., et al., 2003b, A\&A, 411, 565

\bibitem[2002]{spruit} Spruit, H.C., 2002, A\&A, 381, 923 
\end{thebibliography}
\end{document}